%% file: main.tex
\documentclass[a4paper,12pt]{report}
\usepackage{style}
\usepackage{ifpdf}
\def\papertitle{Master Thesis - CHERI in-process isolation}
\def\firstauthor{Sacha Ruchlejmer}

\newif\ifpdf
\ifx\pdfoutput\relax
\else
   \ifcase\pdfoutput
      \pdffalse
   \else
      \pdftrue
\fi
\ifpdf 
  \usepackage[pdftex,
    pdftitle={\papertitle},
    pdfauthor={\firstauthor},
    bookmarksnumbered, 
    pdfstartview=XYZ 
   ]{hyperref}

  \usepackage[pdftex]{graphicx}
  \graphicspath{{./images/}}
  \DeclareGraphicsExtensions{.pdf,.jpeg,.png,.PNG}

  \usepackage[figure,table]{hypcap}

\else 
  \usepackage[dvips,
    bookmarksnumbered, 
    pdfstartview=XYZ 
  ]{hyperref}  

  \usepackage[dvips]{epsfig,graphicx}
  \graphicspath{{./images/}}
  \DeclareGraphicsExtensions{.eps}

  \usepackage[figure,table]{hypcap}
\fi
\usepackage{datetime}
\usepackage{listings}
\usepackage{xcolor}
\usepackage{subcaption}
\usepackage{pifont}
\usepackage{fancyhdr}
\usepackage{amssymb}
\usepackage[utf8]{inputenc}
\usepackage[T1]{fontenc}
\usepackage{geometry}
\usepackage{enumitem}
\usepackage{glossaries}
\usepackage{array}
\usepackage{tabularx}
\usepackage{makecell}
\usepackage{float}
\usepackage{pifont}
\usepackage{xspace}
\usepackage{tcolorbox}

\newcommand{\mydef}[2]{%
  \bigbreak
  \begin{tcolorbox}[boxrule=0.2mm,title={\textbf{#1}}]
    \textit{{#2}}%
  \end{tcolorbox}
  \bigbreak}

\definecolor{codegreen}{rgb}{0,0.6,0}
\definecolor{codegray}{rgb}{0.5,0.5,0.5}
\definecolor{codepurple}{rgb}{0.58,0,0.82}
\definecolor{backcolour}{rgb}{0.95,0.95,0.92}

\newcommand\LSTSize{\fontsize{8.5}{8.2}\selectfont}
\newcommand*\LSTfont{\LSTSize\ttfamily}

\newcommand{\dCOne}{\ding{192}\xspace}
\newcommand{\dCTwo}{\ding{193}\xspace}
\newcommand{\dCThree}{\ding{194}\xspace}
\newcommand{\dCFour}{\ding{195}\xspace}
\newcommand{\dCFive}{\ding{196}\xspace}
\newcommand{\dCSix}{\ding{197}\xspace}
\newcommand{\dCSeven}{\ding{198}\xspace}

\lstdefinestyle{mystyle}{
    language=c,
    backgroundcolor=\color{backcolour},
    commentstyle=\color{codegreen},
    numberstyle=\tiny\color{codegray},
    stringstyle=\color{codepurple},
    basicstyle=\LSTfont,
    breakatwhitespace=false,
    breaklines=true,
    showtabs=false,
    captionpos=b,
    keepspaces=true,
    numbers=left,
    numbersep=5pt,
    showspaces=false,
    showstringspaces=false,
    breakatwhitespace=false,
    tabsize=2,
    escapeinside={\%*}{*)},
    keywordstyle=\color{orange},
    morekeywords={cheri_domain_setup, cheri_domain_enter, cheri_domain_exit, SUCCESSFUL_INITIALIZE, ALREADY_INITIALIZE},
    showtabs=false,
    lineskip=1pt,
}
\lstdefinelanguage{CheriAssembly}{
  morekeywords={stp, str, sub, mov, bl, cbnz, add, ldp, ret}, 
  sensitive=true,
  breakatwhitespace=false,
  breaklines=true,
  showtabs=false,
  keepspaces=true,
  showspaces=false,
  showstringspaces=false,
  morecomment=[l]{//},    
  morestring=[b]"         
}

\lstdefinestyle{CheriStyle}{
  language=CheriAssembly,
  basicstyle=\LSTfont,
  keywordstyle=\color{orange},
  commentstyle=\color{codegreen},
  stringstyle=\color{codepurple},           
  numbers=left,                      
  numberstyle=\tiny\color{codegray},     
  stepnumber=1,                      
  numbersep=5pt,                    
  backgroundcolor=\color{backcolour}, 
  tabsize=2,                         
  captionpos=b,                      
  breaklines=true,                   
  breakatwhitespace=false,           
  showspaces=false,                  
  showstringspaces=false,            
  showtabs=false,                    
}

\title{\papertitle}
\author{\firstauthor}
\date{\monthname[\month] \the\year}

\hypersetup{
    colorlinks,%
    citecolor=black,%
    filecolor=black,%
    linkcolor=black,%
    urlcolor=black
}

\usepackage{placeins}
\usepackage{setspace}

\makeglossaries
\input{glossaries}
\usepackage{cleveref}

\begin{document}

\newgeometry{left=1cm,right=1cm,top=1cm,bottom=2cm} 

\begin{minipage}{0.6\textwidth}
    \includegraphics[width=0.7\textwidth]{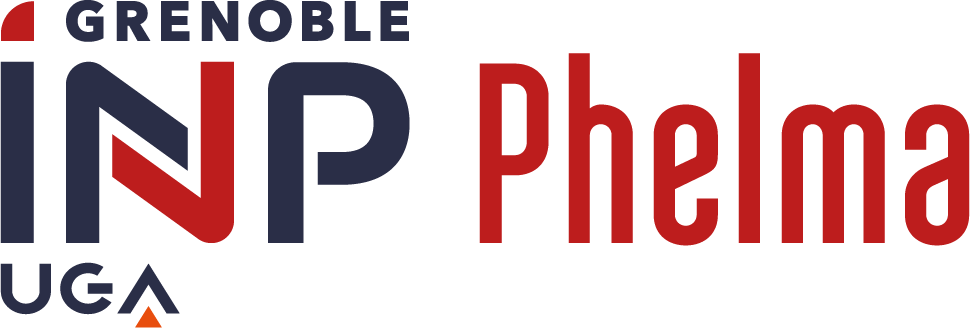}
\end{minipage}
\begin{minipage}{0.29\textwidth}
    \raggedleft
    \includegraphics[width=0.6\textwidth]{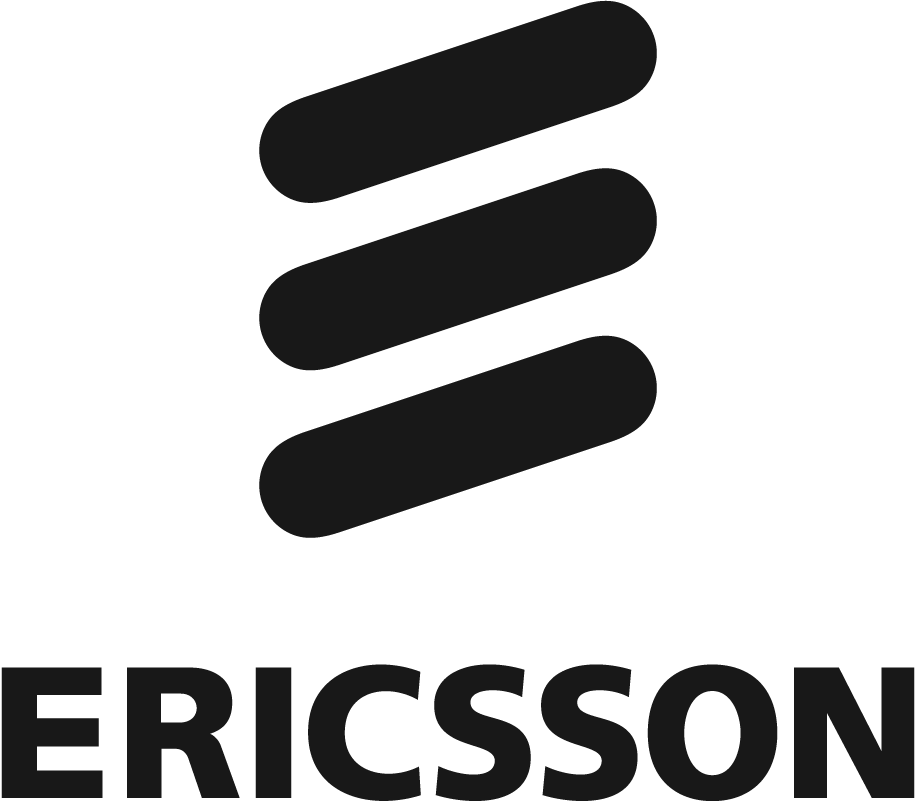}
\end{minipage}
\vspace{1cm}

\begin{center}
    \textbf{\huge Sacha RUCHLEJMER} \\[1cm]
    \textbf{\Large SEOC} \\[0.5cm]
    \textbf{\Large 2024} \\[1cm]
    \textbf{\Large Ericsson} \\[0.5cm]
    \textbf{\Large Torshamnsgatan 21, 164 40 Kista, Sweden} \\[2cm]
    \begin{spacing}{1.7}
    \textbf{\Huge Secure Rewind and Discard on Arm Morello} \\[1cm]
    \end{spacing}
    \textbf{\Large from 21/02/2024 to 07/07/2024} \\[1.5cm]
\end{center}

\begin{flushleft}
    \textbf{Under the supervision of:} \\[0.5cm]
    \begin{itemize}[label=$-$]
        \item Company supervisor: Merve, GÜLMEZ, merve.gulmez@ericsson.com \\[0.35cm]
        \item Phelma Tutor: Cyrille, CHAVET, cyrille.chavet@grenoble-inp.fr
    \end{itemize}
\end{flushleft}

\begin{flushright}
    \textbf{Confidentiality:} $\square$ yes $\boxtimes$ no
\end{flushright}

\begin{flushleft}
    \small
    Ecole nationale \\
    supérieure de physique, \\
    électronique, matériaux \\[0.5cm]
    \textbf{\textcolor[RGB]{230,0,0}{Phelma}} \\
    Bât. Grenoble INP - Minatec \\
    3 Parvis Louis Néel - CS 50257 \\
    F-38016 Grenoble Cedex 01 \\[0.5cm]
    Tél +33 (0)4 56 52 91 00 \\
    Fax +33 (0)4 56 52 91 03 \\[0.5cm]
    \url{http://phelma.grenoble-inp.fr}
\end{flushleft}

\pagenumbering{gobble} 
\restoregeometry

\newpage
\pagenumbering{arabic}
\setcounter{page}{2}
\tableofcontents
\listoffigures
\lstlistoflistings
\listoftables
\printglossaries

\newpage
\pagestyle{fancy}

\section*{Acknowledgments}

I would like to express my sincere gratitude to Merve Gülmez, my thesis supervisor, security researcher at Ericsson, for her invaluable guidance, patience, and encouragement throughout this research. Her expertise and insightful feedback were instrumental in shaping this work. She knew how to give me all the keys I needed to accomplish this work.

\vspace{12pt}
\noindent
I am also thankful to Cyrille Chavet, my school supervisor, for his support, guidance, and for always being attentive to any questions I had.

\vspace{12pt}
\noindent
I am also grateful to Thomas Nyman, a security expert at Ericsson, for his help and support during crucial moments, and for always offering valuable advice.

\vspace{12pt}
\noindent
I would also like to extend my gratitude to Christoph Baumann, a researcher, for his invaluable support, assistance in acquiring the resources I needed, and offering valuable advice.

\vspace{12pt}
\noindent
Special thanks to the other Master's Thesis students and colleagues Sönke and Panagiotis for their stimulating discussions and unwavering support.

\vspace{30pt}
This project is done at the network platform and telecommunication company Ericsson.

\newpage
\glsadd{risc}

\section{Introduction}
Numerous applications are developed using memory-unsafe languages, rendering them susceptible to runtime attacks such as control-flow attacks and data-oriented attacks. These vulnerabilities provide attackers with avenues to gain unauthorized access to programs, by exploiting weaknesses to manipulate and corrupt their behavior. According to a U.S National Security Agency report "The Case for Memory Safe Roadmaps" \cite{memory_roadmap}, two-thirds of reported vulnerabilities in memory-unsafe programming languages still relate to memory issues. Nowadays, there are well-known solutions to mitigate these vulnerabilities like stack canaries \cite{canaries}, but at the end such mitigations terminate the process to prevent attacks exploiting such vulnerabilities from being successful, already corrupted memory will, under normal circumstances prevent the normal operation of the applications. This is especially problematic for service-oriented applications such as web-servers, which must maintain consistent service for all clients even in presence of malicious clients.

State-of-the-art approaches address two related challenges, 1) how to improve the resilience of applications, and 2) how to prevent programs from being exploited by memory-related attacks. 
\textbf{\gls{sdrad}} \cite{gulmez2023} is prior work addressing the first challenge by allowing parts of processes (i.e, sub-processes or routines) to be isolated by creating new, logical protection domains within a conventional process, each with its own stack and heap distinct from the main application stack and heap, and those of other domains. Thanks to this in-process isolation it is possible to discard any domains which memory is corrupted by run-time attacks and going back to a safe anchor point allowing the application to continue running even if the part is corrupted.

\textbf{\gls{cheri}} \cite{CHERI} is prior work addressing the second challenge by extending conventional hardware \glspl{isa} with new architectural features to enable fine-grained memory protection and highly scalable software compartmentalization. It uses the concept of capability-based addressing to store metadata, such as bounds information about pointers that is used to prevent buffer overflows from happening. 

The objective of this thesis is to integrate \gls{sdrad} with \gls{cheri} to introduce resilience into this novel architecture and leverage its security advantages to make it more lightweight and address previous limitations by leveraging the compartmentalization capabilities in CHERI.

In this thesis, \Cref{sec:background} explains memory-related attacks and explores the concept of \gls{cheri}, including an introduction to the secure rewind and discard concept. \Cref{sec:problem} discusses the current limitations of \gls{cheri} and stack canaries, highlighting areas where improvements are needed. \Cref{sec:securedomain} introduces the high level adaptation for secure rewind and discard to the \gls{cheri} architecture (see in \Cref{sec:hle}), explains the implementation of the \gls{cheri}-\gls{sdrad} library (see in \Cref{sec:cherisdrad}), examines use cases and discusses the performance overhead (see in \Cref{sec:usecase}). Lastly, the result of this thesis is discussed in \Cref{sec:conc}.

Secure Rewind and Discard on Arm Morello Master's Thesis artifact will be available at \url{https://github.com/secure-rewind-and-discard/}.

\newpage
\section{Background}\label{sec:background}

\subsection{Memory safety}
In programming, memory safety is a critical concern that should be addressed to prevent undesirable behaviors. Many memory issues arise from developer errors rather than from the language itself. Today, memory-safe languages have demonstrated potential in mitigating a broad range of threats. However, they often come with trade-offs, such as increased resource requirements for executing code and limitations on developers' ability to manage low-level memory. As a result, memory-unsafe languages such as \gls{c} and \gls{c++} remain widely utilized due to their unmatched performance control and suitability for tasks that demand direct memory manipulation~\cite{Nyman2020}.

\paragraph{Memory violations}
Memory violations occur when a program accesses memory in an unintended or unauthorized way, leading to unpredictable behavior or crashes.

\paragraph{Runtime attacks}
Runtime attacks exploit vulnerabilities during program execution, often targeting memory-related weaknesses to inject malicious code or alter the program's intended flow.

\paragraph{Control-flow attacks}
Control-flow attack is one class of runtime attacks. It involves manipulating the sequence of instructions executed by a program, typically through exploiting vulnerabilities in control-flow mechanisms like function pointers or return addresses. Control-flow attacks can be categorized into two main types: code-injection and code-reuse attacks. Code-injection attacks utilize methods like buffer overflows to manipulate the return address of functions, redirecting the program execution to previously introduced malicious code. On the other hand, code-reuse attacks aim to alter the application's behavior by modifying return addresses to unintended existing functions.

\paragraph{Buffer overflows}
Buffer overflows happen when a program writes data beyond the allocated buffer space, potentially overwriting adjacent memory, which can be exploited by attackers to execute arbitrary code or manipulate program behavior. It can be used to overwrite a previous return address with a new one, redirecting execution to a malicious program injected by an adversary. This vulnerability can manifest in various forms; two significant examples include during memory copy operations between variables of different sizes and when an application accepts inputs without verifying if they exceed the allocated buffer size.

\paragraph{Data-oriented attacks}
Data-oriented attack is another class of runtime attacks. It focuses on exploiting vulnerabilities related to how data is processed and accessed within a program, aiming to gain unauthorized access to sensitive information or modify program state. Unlike control-flow attacks, which manipulate the program's execution flow, data-oriented attacks often target weaknesses in how data are handled, such as insecure data storage, inadequate input validation, or insufficient data sanitization. These vulnerabilities can be exploited to steal sensitive information or manipulate program behavior to the attacker's advantage.

\subsection{Capability Hardware Enhanced RISC Instructions}
CHERI (Capability Hardware Enhanced RISC Instructions) \cite{CHERI} represents a collaborative research between SRI International and the University of Cambridge, aimed at reevaluating fundamental design principles in both hardware and software to significantly enhance system security.

\gls{cheri} augments traditional hardware \glspl{isa} with novel architectural elements, enabling fine-grained memory protection and scalable software compartmentalization. These enhancements in memory protection empower memory unsafe programming languages like \gls{c} and \gls{c++} to offer robust, compatible, and efficient defenses against many memory-related attacks such as buffer overflow. Additionally, the scalable compartmentalization capabilities of \gls{cheri} enable fine-grained segmentation of \gls{os} and application code, thereby mitigating the impact of security vulnerabilities in ways not previously feasible with existing architectures.

Notably, \gls{cheri} adopts a hybrid capability architecture, integrating architectural capabilities with conventional \gls{mmu}-based  architectures, microarchitectures, and established software stacks built on virtual memory and \gls{c}/\gls{c++}. This approach allows for gradual and easier integration into existing ecosystems \cite{intro_cheri}.

\subsubsection{Capability-Based addressing}
\gls{cheri} employs a capability-based addressing scheme. The principle is to replace classical pointers with protected objects called capabilities, which will store, in addition to the pointer address, additional information like permissions or the pointer’s intended bounds.

\begin{figure}[H]
\centering
\includegraphics[width=0.45\columnwidth]{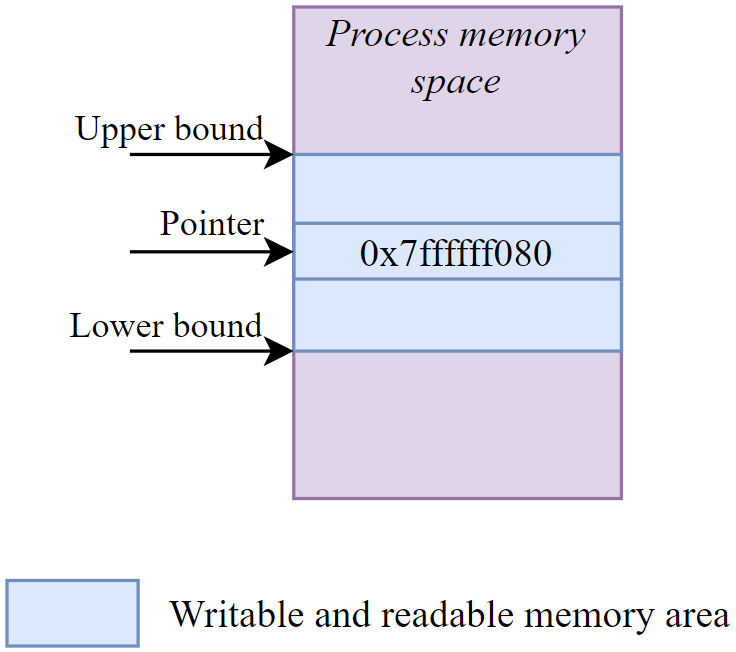}
\caption{Capability representation.}
\label{fig:capabilty_representation}
\end{figure}

As illustrated in Figure \ref{fig:capabilty_representation}, capability-based addressing not only reveals the precise memory location occupied by a capability, but also enables the detection of overflows or overreads before they can occur by explicitly defining the bounds within the pointer's own definition.

\subsubsection{CHERI capability layout}
\begin{figure}[H]
\centering
\includegraphics[width=\columnwidth]{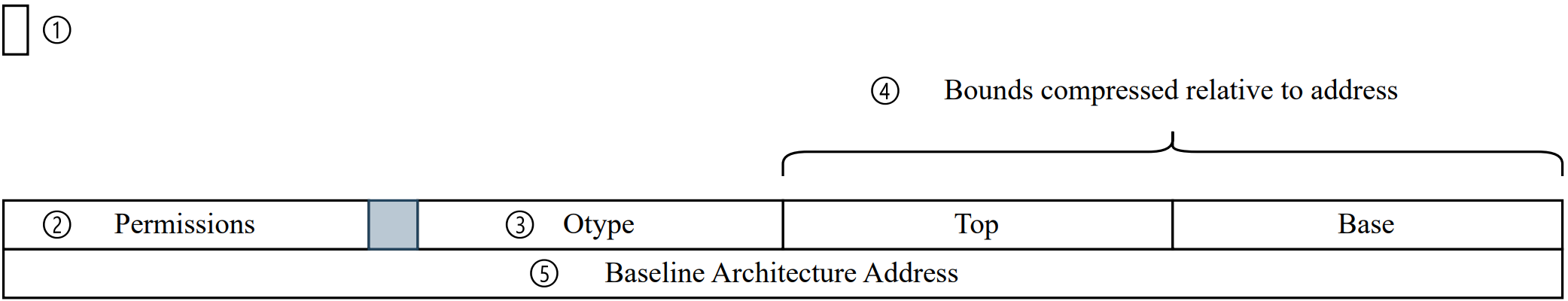}
\caption{CHERI capability layout, adapted from \cite{intro_cheri}.}
\label{fig:CHERI_capability_layout}
\end{figure}

The in-memory layout of a \gls{cheri} capability is shown in Figure \ref{fig:CHERI_capability_layout} where it is depicted as consisting of two layers. The lower layer represents the conventional pointer familiar in traditional programming, while the upper layer includes all the additional information that transforms the pointer into a capability, thus rendering it a protected object. One notable observation is that while a conventional pointer typically occupies 64 bits of memory, the introduction of capabilities necessitates doubling this space to 128 bits.

\noindent
This architecture introduces several enhancements to improve data security and integrity:

\begin{enumerate}
    \item \textbf{Validity Tag} (\ding{172}): Each capability is associated with a 1-bit validity tag, which is maintained automatically in registers and memory. The tag tracks the validity of a capability and ensures that special capability write instructions are needed to create valid capabilities. Regular memory writes, even partial ones, to the memory area of the capability clear the validity tag and invalidate the capability, preventing corrupt capabilities from being dereferenced.
    
    \item \textbf{Permissions} (\ding{173}): The permissions mask controls how the capability can be used, such as by restricting loading (read) and storing (write) of data and/or capabilities, or by prohibiting instruction fetch (execute). Every load, store, or instruction fetch in a \gls{cheri}-enabled microprocessor architecture must be authorized by an architectural capability with the corresponding permissions.
    
    \item \textbf{Object Type (Otype)} (\ding{174}): Object types allow multiple capabilities to be associated with each other, facilitating software compartmentalization where only a specific set of capabilities can be used within a logically isolated compartment.
    
    \item \textbf{Bounds} (\ding{175}): The lower and upper bounds describe the portion of the address space to which the capability authorizes loads, stores, and/or instruction fetches, depending on the permissions the capability grants.
    
    \item \textbf{Baseline Architecture Address} (\ding{176}): A conventional pointer in the underlying \gls{isa} native format.
\end{enumerate}

\gls{cheri} implements two modes of operation. The first, called hybrid mode, allows conventional pointers and \gls{cheri} capabilities to coexist and be used independently. In most cases, this enables programs that were not originally developed for a \gls{cheri} architecture to continue functioning on \gls{cheri}-enabled hardware. The second mode, known as pure-capability (often referred to as purecap), is the most secure mode of operation. In this mode, pointers are completely replaced by capabilities, resulting in the most secure programs utilizing \gls{cheri} technology.

\subsubsection{Arm Morello Board}
The Arm Morello development board (referred to as subsequently as simply “the Morello board") \cite{morello_board} is an industrial demonstrator of a capability architecture developed by Arm, featuring a prototype System-on-Chip (SoC). This board incorporates a \gls{cheri}-extended ARMv8-A processor, GPU, peripherals, and a memory subsystem. The Morello board allows for hardware and software to be tested in real-world conditions, enabling the evaluation of \gls{cheri}'s viability and performance impact. Its primary objectives are to facilitate industrial evaluation of \gls{cheri} hardware and software concepts, gather evidence for potential adoption, and support ongoing research and development. By integrating \gls{cheri} into a widely deployed, real-world architecture with a high-end, mature processor design and a robust software ecosystem, the Morello board aims to advance the practical application of these technologies.

The Morello board runs an adapted version of the FreeBSD \gls{os} called CHERIBSD.
The experimental work described in this thesis has been conducted using a Morello board on loan to Ericsson to obtain firsthand insight into the performance metrics, rather than relying solely on simulated data, which could be influenced by the underlying hardware support.

\subsection{Secure Domain Rewind and Discard}
\gls{sdrad} (Secure Domain Rewind and Discard)\cite{gulmez2023} project aims to improve the resilience of applications against run-time attacks. Indeed, today's mitigation techniques against run-time attacks terminate the application when they detect an attack. However, terminating the application in response to an attack is disruptive to service-oriented applications that service many independent clients simultaneously. If 
 the service process is terminated all clients being serviced will lose their connections because of one attack. Motivated by this, alternative approaches are being explored, with \gls{sdrad} being one of them. The main idea behind \gls{sdrad} is to isolate different parts of a program from each other in separate domains using in-process isolation. This domain mechanism allows the code inside it to be executed in a different part of memory than the process's own. When a new domain is created, it initializes with its own stack and heap. Additionally, to prevent access to unwanted memory spaces, such as those of other domains, an isolation mechanism, such as \glspl{mpk} is used to check if the rights to use this memory space are present. This isolation prevents domains from accessing each other's memory, which stops any domain affected by a run-time attack from corrupting memory belonging to another domain.

The initial \gls{sdrad} prototype implementation targeted the 64-bit x86 architecture and utilized \gls{mpk}, a technology developed by Intel and introduced in the "Skylake" microarchitecture \cite{mpk}. \gls{mpk} is a hardware feature that provides memory protection at the page granularity. The main purpose of \gls{mpk} is to allow software developers to define memory regions with specific protection keys. This enables fine-grained control over memory access without the need to rely on \gls{kernel}-enforced access control for memory pages, which is required for regular processes. Such reliance adds a costly context switch every time the execution context changes from one protection domain to another.
In practice, MPK works by associating a protection key with each memory page. The permissions for the keys are stored in a special register called the protection key register, and the keys are added to the page table. The page table is a data structure used by the \gls{os} to map virtual addresses to physical addresses in memory. A program can specify a protection key when accessing a memory page, and the processor checks if the provided key matches the key associated with that page without involving the \gls{kernel}. If the keys match, memory access is allowed; otherwise, an exception is triggered.

In summary, \gls{sdrad} provides an in-process-based solution that allows compartmentalizing the application into distinct domains, where each domain operates independently and can be discarded if its memory has been corrupted, and guarantees that memory belonging to other domains are unaffected.

\newpage
\section{Problem Statement: Limitation with CHERI and other defenses}\label{sec:problem}

\begin{figure}[ht]
\centering
\includegraphics[width=0.6\columnwidth]{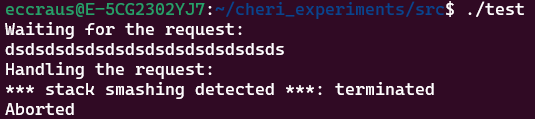}
\caption{Handling buffer overflows with canaries.}
\label{fig:stack_canaries}
\end{figure}

\begin{figure}[H]
\centering
\includegraphics[width=0.75\columnwidth]{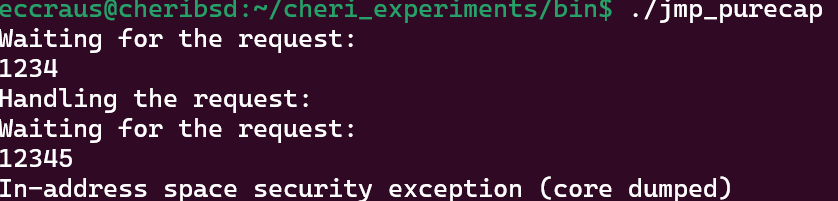}
\caption{Handling buffer overflows with CHERI.}
\label{fig:result_cheri_overflow}
\end{figure}

The limitations of current mitigation techniques and the defense provided by \gls{cheri} is demonstrated in Listing \ref{lst:test_overflow}, \ref{fig:stack_canaries} and \ref{fig:result_cheri_overflow} using a small program that prompts a user to enter characters into the program buffer.

\begin{lstlisting}[caption={Example of unsafe code.}, label={lst:test_overflow}]
void get_request(){
    char buff[5];
    printf("Waiting for the request:\n");
    scanf("%s",buff);
    printf("Handling the request\n");
}

int main(void){
    int i = 0;
    for(i = 0; i < 5; i++){
        get_request();
    }

    return 0;
}
\end{lstlisting}

Listing \ref{lst:test_overflow} shows an example of a program that uses an unsafe \gls{c} function, \texttt{scanf()}, to read user input. The \texttt{scanf()} function is implemented in the correct way, but it's unsafe because it has no robust input validation. In classic \gls{c}, entering a string larger than the buffer size does not generate an error. Although there are existing solutions such as stack canaries \cite{canaries} that can detect such attacks, as shown in Figure \ref{fig:stack_canaries}.

As illustrated in Figure \ref{fig:stack_canaries}, the stack canaries are able to detect when a user attempts to input too many characters into the buffer. A stack canary is a random value placed after local variables on the stack. Before and after potentially unsafe operations (e.g. updating local variables), the program checks the integrity of the canary value. If the program detects that the canary was overwritten, it indicates that the input exceeds the size allocated for the variable, resulting in a buffer overflow. As a result, because already corrupt memory cannot be recovered, the application must be terminated completely upon detecting the buffer overflow.

However, with the use of \gls{cheri} capabilities and their metadata describing the bounds of the object a pointer refers to, such as the buffer buff in Listing \ref{lst:test_overflow}, it becomes possible to detect at the hardware level if the string intended for the buffer exceeds the allocated size. As illustrated in Figure \ref{fig:result_cheri_overflow}, \gls{cheri} raises an exception if the string exceeds the specified bounds of the capability.

\gls{cheri} is also able to detect attempts to perform a buffer overflow. In this case the detection occurs \textit{before} any memory corruption takes place. Therefore, using \gls{cheri} provides better defense because the memory remains intact. However, it does not address the issue of resilience, as the program is immediately stopped upon detecting a memory violation.

\vspace{12pt}
 
\mydef{The Problem Statement}{How can the memory-safety guarantees provided by the \gls{cheri} architecture be combined with Secure Rewind and Discard to improve software resilience against run-time attacks?}

\newpage
\section{Secure Domain Rewind and Discard on Arm Morello Board}\label{sec:securedomain}

The goal of this thesis is to adapt the Secure Domain Rewind and Discard to the \gls{cheri} architecture, allowing vulnerable applications to be recovered upon detecting an attack. \gls{cheri} is capable of detecting buffer overflows before memory corruption occurs. This method eliminates the need to compartmentalize the stack when creating new domains, thereby allowing for a lighter-weight \gls{sdrad} design. 
The design requires:
\begin{enumerate}
    \item Creating return points that splits the application into distinct crash-resistant domains.
    \item Leveraging \gls{cheri}'s hardware capabilities to detect attacks.
    \item Heap compartmentalization to keep track of domain heap allocation.
\end{enumerate}

\subsection{High-level idea}\label{sec:hle}
The objective of adapting \gls{sdrad} to the \gls{cheri} architecture is to provide a solution that allows for crash-resistance and resilience against runtime attacks that exploit memory vulnerabilities but also to make it lighter and with better performance compared to the prototype \gls{sdrad} implementation targeting 64-bit x86. 

\begin{figure}[H]
\centering
\includegraphics[width=0.99\columnwidth]{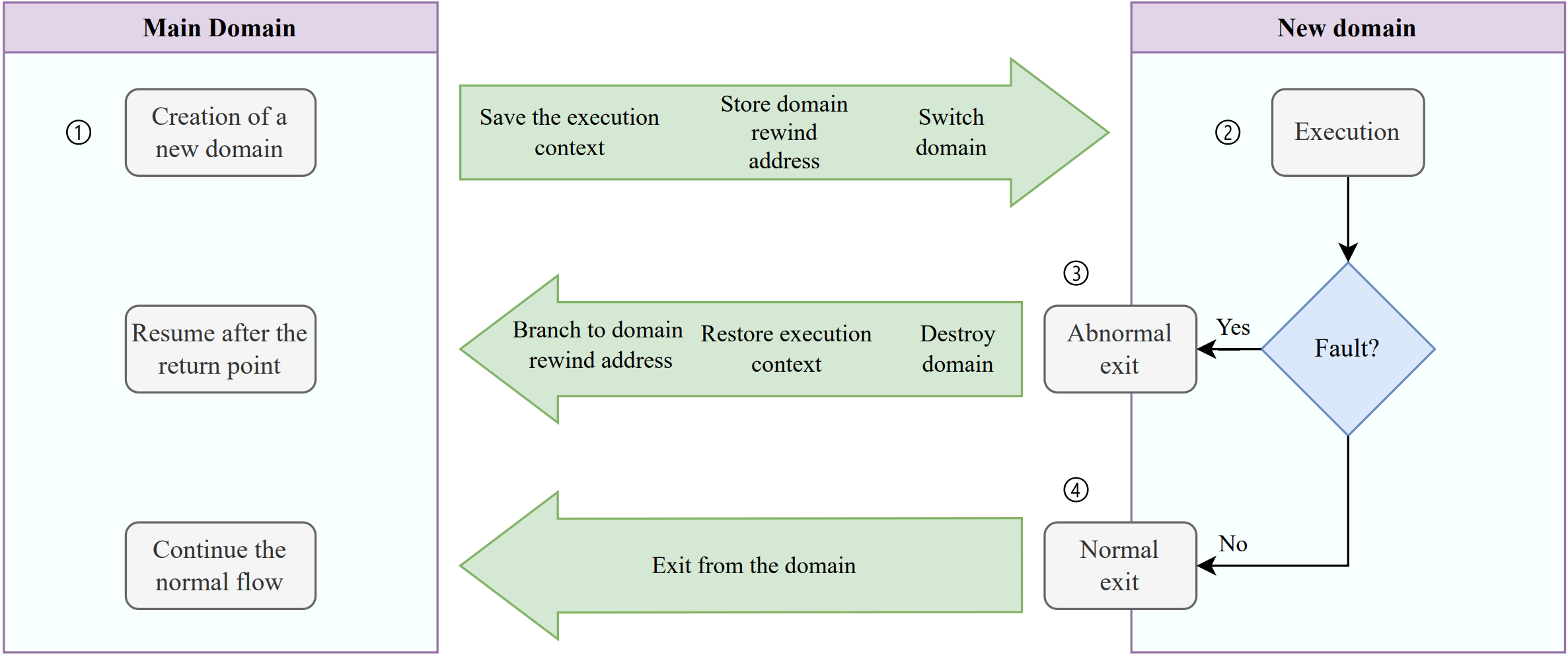}
\caption{High-level idea of SDRaD for CHERI, adapted from \cite{gulmez2023}.}
\label{fig:high_level_idea}
\end{figure}
Figure \ref{fig:high_level_idea} illustrates the high-level idea of \gls{sdrad}. If a particular section of code in a program is at risk and needs to be isolated with the possibility of being rewound, it can be placed into a new domain. Initially, the application runs within the main domain. Before executing the risky code, a new domain is created \ding{172} by saving the execution context (i.e., the program's status) and the current memory address to establish a checkpoint for potential rollback. The program then enters the newly created domain. \ding{173} Within this domain, two scenarios are possible. If memory corruption is detected, the domain exits prematurely with an abnormal exit \ding{174}, destroys the faulty domain, restores the saved execution context, and returns to the main domain, indicating the previously created domain exited abnormally. The caller is expected to handle abnormal exits, similar to code that might throw exceptions. If no memory corruption is detected, the domain exits normally \ding{175}, and the program continues its usual flow. In that case, this domain still exists and can be used again later. This approach ensures that risky code can be isolated and rewound from if necessary, without affecting the main domain, thus enhancing the program's resilience.

\subsection{CHERI-SDRaD}\label{sec:cherisdrad}
The \gls{cheri}-\gls{sdrad} \gls{c} library was developed to adapt \gls{sdrad} for the \gls{cheri} architecture using approximately 1.2k LoC of \gls{c} code and 20 lines of Arm assembly code. This library provides an \gls{api} that allows developers to integrate secure rewind and discard mechanisms into their applications. The following section will describe the design and implementation of the \gls{cheri}-\gls{sdrad} library.

\subsubsection{Domain Manager}
To manage  the domains created within the application, a global manager was introduced, as illustrated in Listing \ref{lst:manager}. The Domain Manager is defined as a global variable so that it can be accessed from anywhere in the code. It has two members: the first one, \texttt{active\_domain} \ding{172}, indicates the current domain, and the second, \texttt{domain\_info} \ding{173}, stores the informations about currently existing domains. One limitation of the Domain Manager is that domain information is stored in an array with a static size (here 16). The maximum number of domains supported by the \gls{cheri}-\gls{sdrad} library must be defined at compilation time. This limitation could be addressed in future work by dynamically defining the domains, such as using a linked list. However, one advantage of the array-backed domain information storage is
that information of any domain can be accessed in constant-time.

\newpage
\begin{lstlisting}[caption={CHERI-SDRaD manager.}, label={lst:manager}]
#define NUMBER_MAX_DOMAIN 15
enum State{UNINIT, INIT};

typedef struct _return_reg_type_s {
    void *c29;
    void *c30;
}return_reg_type_s;

typedef struct _domain_info_s {
    jmp_buf env;
    return_reg_type_s return_address;
    tlsf_t tlsf;
    uint32_t parent_udi; %*\dCThree*)
    enum State domain_init;
    enum State heap_init;
} domain_info_s;

typedef struct _global_manager_s{
    uint32_t active_domain; %*\dCOne*)
    domain_info_s domain_info[NUMBER_MAX_DOMAIN+1]; %*\dCTwo*)
} global_manager_s;
\end{lstlisting}

Furthermore, the \gls{cheri}-\gls{sdrad} manager includes a \texttt{parent\_udi} \ding{174} in the \texttt{domain\_info} section. This means that domain nesting, i.e., having one domain inside another, is possible recursively within this implementation.

\subsubsection{CHERI-SDRaD API}
Developers can use the \gls{api} specified in Table \ref{tab:api_function} to improve resilience inside their application.
\begin{table}[H]
    \centering
    \resizebox{0.8\textwidth}{!}{
    \begin{tabularx}{\textwidth}{|l|c|p{0.4323\textwidth}|}
      \hline
      \gls{api} function name & arguments & description \\
      \hline
      \ding{172} cheri\_domain\_setup() & udi & \makecell[l]{Create a new domain with \\ the specified udi}  \\
      \hline
      \ding{173} cheri\_domain\_enter() & udi & \makecell[l]{Enter inside an already created  \\ domain with its udi}  \\
      \hline
      \ding{174} cheri\_domain\_exit() & - & \makecell[l]{Exit from the current \\ domain} \\
      \hline
    \end{tabularx}
    }
    \caption{List of API functions.}
    \label{tab:api_function}
\end{table}

Domains are initialized by invoking \texttt{cheri\_domain\_setup} \ding{172}. This function creates a new domain with a unique \gls{udi} and prepares it for use by saving the execution context and the return address to the Domain Manager, and completes the initialization process. Upon initialization, this call provides a return value: a positive value indicates successful initialization or that the domain with the specified \gls{udi} is already initialized, while a negative value signifies an out-of-bounds \gls{udi}.
Once initialized, a domain can be entered using \texttt{cheri\_domain\_enter} \ding{173}. This operation returns an error code if the specified \gls{udi} is out of bounds or not associated with any existing domain. Conversely, it provides a success code if the \gls{udi} is correct, enabling entry into the specified domain.
When a domain is no longer needed, it can be exited by calling \texttt{cheri\_domain\_exit} \ding{174}, returning to its parent domain.

Listing \ref{lst:example_with_api} illustrates how the previous unsafe code can be encapsulated into a domain to ensure resilience in case of any corruption.

\begin{lstlisting}[caption={Unsafe code encapsulated into a new domain.}, label={lst:example_with_api}]
int main(void){
    int i = 0;
    int err;
    int uid = 1;
    for(i = 0; i < 5; i++){
        %*\dCFive*) err = cheri_domain_setup(uid); %*{\dCOne}*)
        if(err == SUCCESSFUL_INITIALIZE || err == ALREADY_INITIALIZE){
            cheri_domain_enter(uid); %*\dCTwo*)
            get_request(); %*\dCFour*)
            cheri_domain_exit(); %*\dCThree*)
    		} else {
            printf("Bad input!\n");
        }
    }

    return 0;
}
\end{lstlisting}

By using the \gls{api}, the \texttt{get\_request} \ding{175} function now operates in a different domain from the main application. The \texttt{err} variable \ding{177} captures the return value sent by \texttt{cheri\_domain\_setup}. If this value indicates successful initialization, \texttt{err} will have a positive value, signifying that the domain is correctly initialized. In this case, the \textit{if} condition is met, allowing entry into the domain to start computing the unsafe function.If no bad behavior is detected during the execution of the unsafe function, the domain is exited by calling \texttt{cheri\_domain\_exit}, and the for loop continues from the next iteration. However, if the user provides input that would overflow the destination buffer, the invalid buffer operation is detected by \gls{cheri}, but the signal indicating that an invalid memory access was about to happen is captured by a signal handler in \gls{cheri}-\gls{sdrad}, which causes a abnormal domain exit. This is indicated to the caller via a false return value stored in \texttt{err}. Consequently, the \textit{if} condition is no longer satisfied, causing the program to enter the \textit{else} block, which warns the user about the invalid input.

\begin{figure}[H]
    \centering
    \includegraphics[width=0.8\textwidth]{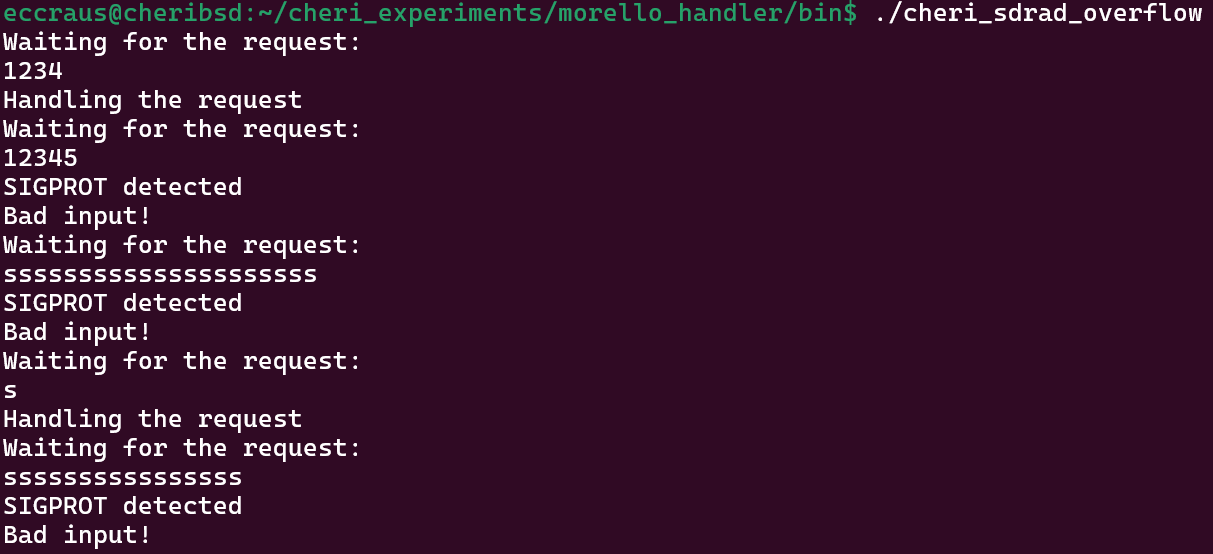}
    \caption{Handling buffer overflows with CHERI-SDRaD.}
    \label{fig:result_example_with_api}
\end{figure}

Figure \ref{fig:result_example_with_api} shows, using the example application from Listing \ref{lst:example_with_api}, how using \gls{cheri}-\gls{sdrad} handles buffer overflows in a domain. When a user submits an input that is too large, the routine is interrupted, causing the domain to exit without completing the execution of the code inside. As a result, when encountering a bad input, the request is not processed further. However, a too large input does not stop the application; it continues running until completion. Therefore, the resilience of this application has been improved by using \gls{cheri}-\gls{sdrad}. 

\subsubsection{Saving the execution context}
To efficiently handle abnormal domain exits and safely rewind to the main domain. When establishing a new domain, three pieces of information must be preserved: the new \gls{udi}, which identifies the newly created domain; the domain rewind address, which instructs the system on where is the entry point to which execution is rewound if the domain needs to be abnormally exited; and the execution context, encompassing all the elements that are required to allow a program to resume its operation at the specific point and state before the domain was created. This includes the stack pointer, the program counter, the link register, and all general-purpose registers. The \gls{udi} and domain rewind address are straightforward to retrieve and store as they can be directly accessed within the code. However, saving the execution context and being able to return to it requires using a special \gls{c} standard library call, \textbf{setjmp}. The setjmp \gls{api} library provides two functions: \texttt{setjmp()}, which saves the execution context, and \texttt{longjmp()}, which allows to return to a previously saved execution context, effectively restoring the state to where \texttt{setjmp()} was called. The \texttt{setjmp()} function takes one argument of type \texttt{jmp\_buf}, which is a special type provided by the \gls{api}. When called, \texttt{setjmp()} stores the execution context in the \texttt{jmp\_buf} variable given as an argument. 
The \texttt{longjmp()} function takes two arguments: the first one is a \texttt{jmp\_buf} variable, which represents the execution context to be restored, and the second one is a strictly positive value which specified the return code code for \texttt{setjmp()}.

\begin{figure}[H]
    \centering
    \begin{subfigure}[b]{0.47\textwidth}
        \centering
        \includegraphics[width=0.5\columnwidth]{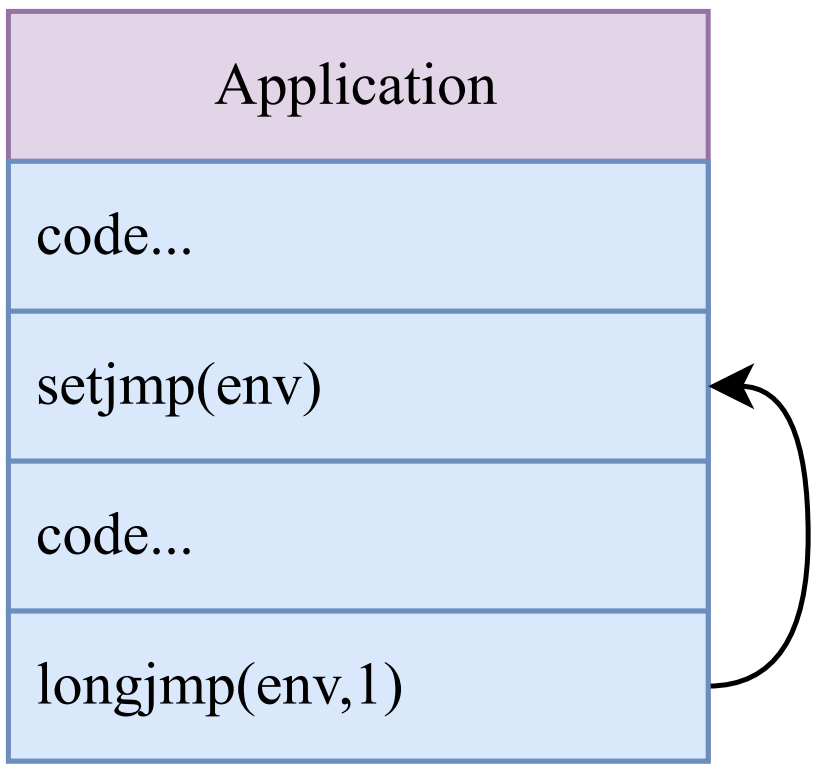}

        \caption{Normal setjmp.}
        \label{fig:setjmp_classic}
    \end{subfigure}
    \hfill
    \begin{subfigure}[b]{0.47\textwidth}
        \centering
        \includegraphics[width=\columnwidth]{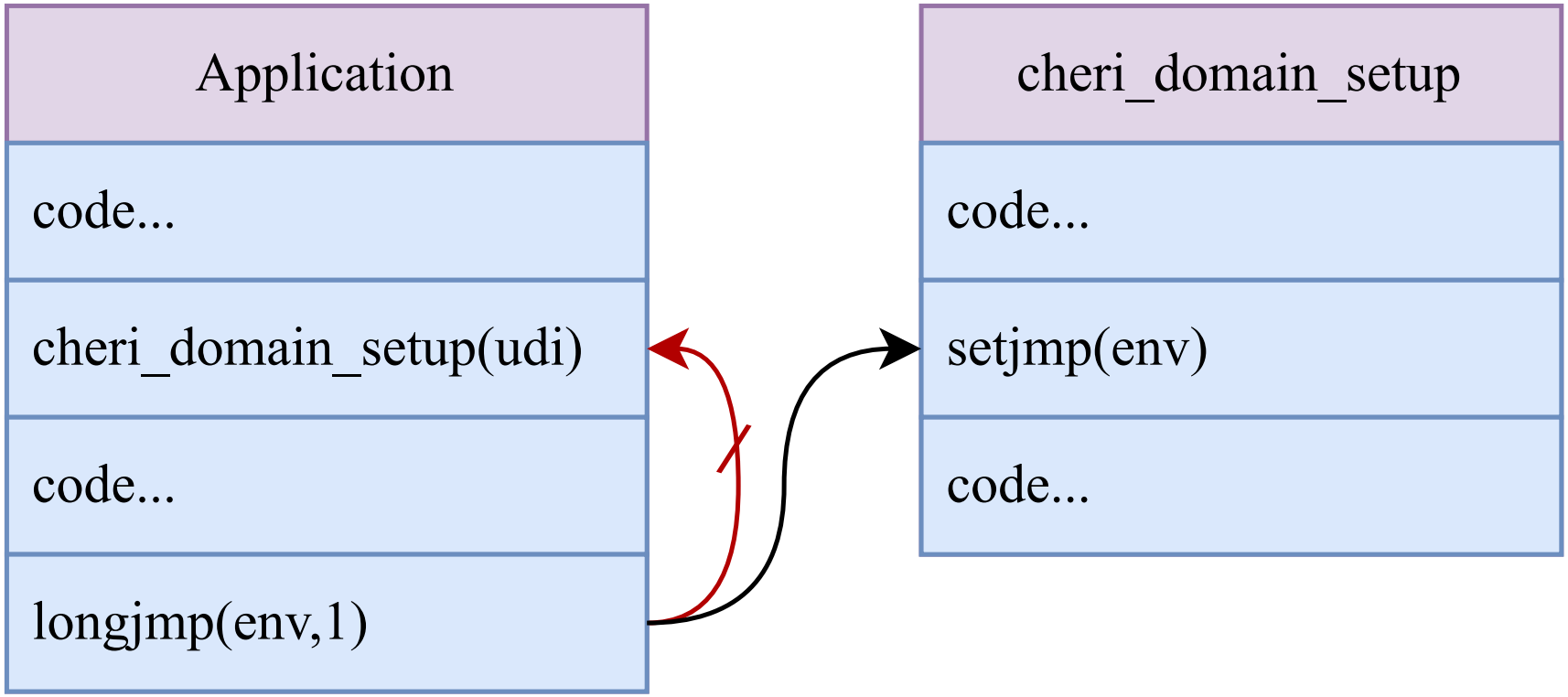}
        \caption{Setjmp inside API.}
        \label{fig:setjmp_api}
    \end{subfigure}
    \caption{Problem with setjmp in the API.}
    \label{fig:setjmp_trick}
\end{figure}

Normally, \texttt{setjmp()} would be used to save the execution context, but with the \gls{api} adding an intermediate function layer, the execution context cannot be save as usual. The problem is shown in Figure \ref{fig:setjmp_trick}. 

Invoking \texttt{longjmp()} with a stored context created in a function that has already returned results in undefined behavior. The solution is to devise a trampoline directly in assembly code. This trampoline ensures the context stored in the \texttt{jmp\_buf} when \texttt{cheri\_domain\_setup} is invoked matches the context of the caller.
By using an \gls{assembly} function and directly manipulating memory, this trampoline tricks the system into believing that the call to \texttt{setjmp()} occurred within the same function that initiated the \gls{api} call. After the context has been saved, the trampoline invokes the code that creates the necessary run-time structures for the new domain, as shown in Listing \ref{lst:assembly}.

\begin{lstlisting}[style=CheriStyle, caption={Assembly function that stores initialization metadata.}, label={lst:assembly}]
cheri_domain_setup:
    stp     c29, c30, [csp, #-32]! %*\dCOne*)     // store the rewind address on the stack
    str     c0, [csp, #-16]! %*\dCTwo*)           // store the udi on the stack
    sub     csp, csp, #512              // store some space for the env variable on the stack

    mov     c0, csp %*\dCThree*)                    // give the environnement variable to setjmp
    bl      setjmp@PLT
    cbnz    w0,.ljmp                    // if non 0 go to .ljmp
    mov     c0, csp %*\dCFour*)
    bl      cheri_domain_init@PLT

    add     csp, csp, #528              // restore the stack pointer
    ldp     c29, c30, [csp], #32   %*\dCFive*)       // restore the return address and the stack pointer
    ret     c30   %*\dCSix*)

.ljmp:
    bl      cheri_domain_destroy@PLT
    ldp     c29, c30, [c0]              // load the return address
    add     csp, csp, #0x230            // restore the stack pointer
    ret     c30   %*\dCSeven*)
\end{lstlisting}


The \gls{assembly} function, \texttt{cheri\_domain\_setup()}, illustrated in Listing \ref{lst:assembly} is the one used by the \gls{api} to initiate a new domain. As mentioned earlier, to create a new domain, three pieces of information(\gls{udi}, execution context and return address) need to be saved by employing a trick within the application. 

In \gls{assembly}, a limited number of registers are available for data manipulation. One way to overcome this limitation is by saving register values onto the stack, which is a memory area functioning as a \gls{lifo} structure. The stack pointer register (here \texttt{csp}) points to the last empty location in this memory area. 

When a function is called, the address of the call is automatically stored in the \texttt{c30} register. The first step of this trick is to decrement the stack pointer to allocate space for storing the domain rewind address, the \gls{udi} of the new domain and the execution context \ding{172}. The second line of the \gls{assembly} code stores the value of \texttt{c0} on the stack because when an argument is passed to a function, it is stored in the \texttt{c0} register for later use \ding{173}.

Note that, Arm Morello architecture~\cite{arm_morello_manual} defines that if branch with a link instruction generates a sealed capability in c30, to unseal c30 later it must be used in a valid unseal, operation, such as ret in \ding{197}, \ding{198}. The sealed c30 and frame pointer in c29 need to be saved and restored using the store capability pair (STP) \ding{192} and load capability pair (LDP) \ding{196} instructions to avoid invalidating their tag bits. 

The next step is to store the current stack pointer into \texttt{c0}. Recall that the stack pointer holds the address of the newly allocated \texttt{jmp\_buf} allocated on the stack. By storing that address into \texttt{c0} the address is passed as argument to \texttt{setjmp()} that will save the execution context into the allocated space \ding{174}. The \texttt{setjmp()} function returns 0 if the execution context was saved successfully. A non-zero return value for \texttt{setjmp()} indicates it was executed as a result of a \texttt{longjmp()} call that restored the previously saved execution context. The latter case corresponds to an abnormal domain exit, in which case the trampoline code destroys the saved information for the faulting domain by calling the internal \texttt{cheri\_domain\_destroy()} functions and returns from the trampoline by restoring the saved return address from \texttt{jmp\_buf}.

The final part of the trampoline \ding{175} is to store the stack pointer into \texttt{c0} so that \texttt{cheri\_domain\_init()} can access it as an argument and store these variables. Listing \ref{lst:cheri_domain_init} illustrates this function. The \texttt{cheri\_domain\_init()} function will store the execution context, the \gls{udi}, and the rewind address from the stack into variables \ding{172}. These variables will eventually be stored in the \gls{cheri}-\gls{sdrad} Manager if the initialisation is successful. The order of the fields in the \texttt{cheri\_init\_stack\_s} struct is important because, since the stack operates on a \gls{lifo} basis, the last argument stored in the assembly code will be the first one stored in the \texttt{cheri\_init\_stack\_s} and vice-versa.
The next steps are to check if this domain can be created \ding{173}, and if so, to verify if it already exists \ding{174}.
If initialization of the domain is successful and it does not already exist, then the domain is initialized \ding{175}. The parent domain that invoked the creation of this new domain is saved \ding{176}, along with the rewind address \ding{177} and the execution context \ding{178}, stored in the \texttt{domain\_info} field of the \gls{cheri}-\gls{sdrad} Manager.

\begin{lstlisting}[caption={cheri\_domain\_init function.}, label={lst:cheri_domain_init}]
struct cheri_init_stack_s{
    jmp_buf env;
    uint64_t udi; 
    return_reg_type_s return_address;
};

int cheri_domain_init(void *base_address){
    struct cheri_init_stack_s *cis_ptr;
    cis_ptr = (struct cheri_init_stack_s *)base_address; %*\dCOne*)
    long udi = cis_ptr->udi;
    global_manager_s *gm_ptr = &cheri_sdrad_manager;

    if(udi > (NUMBER_MAX_DOMAIN) || udi < 1){ %*\dCTwo*)

        printf("invalid udi, you should choose one between 1 and %d\n",NUMBER_MAX_DOMAIN);
        return UDI_OUT_OF_BOUNDS;
    }

    if(gm_ptr->domain_info[udi].domain_init == INIT){ %*\dCThree*)
        printf("This domain is already initialised\n");
        return ALREADY_INITIALIZE;
    }
    gm_ptr->domain_info[udi].domain_init = INIT; %*\dCFour*)
    
    gm_ptr->domain_info[udi].parent_udi = manager.active_domain; %*\dCFive*)
    
    gm_ptr->domain_info[udi]. return_address = cis_ptr->return_address; %*\dCSix*)
    memcpy(gm_ptr->domain_info[udi].env, cis_ptr->env, sizeof(jmp_buf)); %*\dCSeven*)

    return SUCCESSFUL_INITIALIZE;
}
\end{lstlisting}

\subsubsection{CHERI protection violation handler}
The second step in implementing resilience is to modify the application's behavior to prevent it from crashing. An application crash occurs when a program encounters an error or a set of conditions that it cannot handle, leading to an abrupt termination of its operation.
\gls{sdrad} works on both Intel and AMD architectures by detecting the SIGSEGV signal that can be attributed to \gls{mpk}-related access faults, such domain violation, or SIGABRT sent as a result of a failed stack canary check. However, on the Morello board, a \gls{cheri} capability fault is reported by the \gls{os} to the application via a ”protection violation fault” (SIGPROT) signal. Therefore, the \gls{cheri}-\gls{sdrad} signal handler had to be adjusted accordingly, as illustrated in Listing \ref{lst:signal_handler}.
 
\begin{lstlisting}[caption={Signal handler code.}, label={lst:signal_handler}]
__attribute__((constructor)) %*{\dCOne}*)
void cheri_setup_signal_handler()
{
    struct sigaction sa;
    sa.sa_flags =  SA_SIGINFO;
    sa.sa_handler = cheri_signal_handler;
    sigemptyset(&sa.sa_mask);
    if (sigaction(SIGPROT, &sa, NULL) == -1) {
        printf("sigaction");
    }
}

void cheri_signal_handler(int signum)
{
    global_manager_s *gm_ptr = &cheri_sdrad_manager;
    domain_info_s *di_ptr = &(gm_ptr->domain_info[gm_ptr->active_domain]);
    int udi = gm_ptr->active_domain;

    if(signum == 34){
        if(udi != 0){
            printf("SIGPROT detected\n");
            
            longjmp(di_ptr->env,14);
        }
        else{
            exit();
        }
    }else{
        printf("signum: %d\n",signum);
    }
}
\end{lstlisting}

The \textbf{constructor} attribute\footnote{\url{https://clang.llvm.org/docs/AttributeReference.html\#constructor}} is a Clang attribute that allows a function to run before the main execution start. This attribute is associated with the\texttt{cheri\_setup\_signal\_handler} function \ding{172}, ensuring that this function is executed before the main function. The \texttt{cheri\_setup\_signal\_handler} will change the behavior of the \gls{os} signal delivery delivering mechanism to use a custom handler when delivering \texttt{SIGPROT}. The new \texttt{SIGPROT} handler, \texttt{cheri\_signal\_handler()},verifies if a \texttt{SIGPROT} signal has indeed been detected. If so, it also checks whether the active domain is not the main one, closing the application if it is. Finally, if all conditions are met, it uses the \texttt{longjmp()} function to restore the execution context saved prior to creating this domain, allowing the program to resume from that point.

\subsubsection{Heap management}
Implementing heap management for \gls{cheri}-\gls{sdrad} allows to maintain separate heaps for different domains and ensures that allocations occur in the appropriate heap. This is important to make sure that allocations that occur in domains that exit abnormally can be freed without leaking memory.
\paragraph{Heap allocator}
Creating an isolated domain requires generating an isolated heap for each domain. An attempt was made initially to adapt a compartmentalizing allocator from the Cambridge \glspl{cheri} project \cite{Heap_compartment} to \gls{cheri}-\gls{sdrad}. The compartmentalizing allocator was designed for software operating in hybrid-capability mode to compartmentalize its heap into smaller, isolated areas. However, it was found during the adaptation of the allocator for purecap mode that the proof-of-concept compartmentalizing allocator did not support freeing individual allocations, only full compartments. This posed an issue for integration into \gls{cheri}-\gls{sdrad} since an allocator was needed that could be used as a drop-in replacement for POSIX \texttt{malloc()} and \texttt{free()} on code paths exiting domains normally.
The original \gls{sdrad} prototype employs \gls{tlsf} allocator that allows allocations to be directed to distinct ”memory pools”. For a meaningful comparison between \gls{cheri}-\gls{sdrad} and the 46-bit x86 \gls{sdrad} prototype, it would be advantageous to adapt \gls{tlsf} as well. However, that meant porting the \gls{tlsf} allocator to \gls{cheri}.

\subsubsection{TLSF}
The Two-Level Segregated Fit (\gls{tlsf}) memory allocator is designed for efficiency by reducing fragmentation and optimizing memory utilization. \gls{tlsf} divides memory into segregated blocks based on their size, establishing separate pools to cater to different ranges of block sizes. This segmentation ensures that memory blocks are allocated with minimal wastage and fragmentation, as blocks of similar sizes are grouped together. 
\gls{tlsf} employs a two-level structure, consisting of broad and fine levels of segregation. At the broad level, memory is divided into larger chunks, while at the fine level, these chunks are further subdivided into smaller blocks. This hierarchical organization facilitates efficient searching and allocation of memory blocks, significantly reducing overhead and improving allocation speed.
\gls{tlsf} can split or merge memory blocks as needed to accommodate varying allocation sizes, further minimizing fragmentation and improving memory utilization.
Moreover, \gls{tlsf} is designed with low overhead in mind, both in terms of memory usage and processing time. This makes it particularly suitable for deployment in resource-constrained environments such as embedded systems and real-time operating systems, where efficient memory utilization is critical for optimal performance.

\paragraph{Implementation}
\gls{sdrad} \gls{tlsf} is based on an open-source project developed by Matt Conte \cite{tlsf}. The \gls{tlsf} implementation was ported to \gls{cheri} by accommodating the increased size of the in-memory representation of \gls{cheri} capabilities compared to the baseline architecture pointers where necessary. For example, the \texttt{ALIGN\_SIZE\_LOG2} variable was changed from 3 to 4 to ensure that the addresses of memory allocations made by \gls{tlsf} are aligned to 16 bytes instead of 8 bytes. Moreover, the values of \texttt{block\_header\_overhead} and \texttt{block\_start\_offset} need to be updated by doubling \texttt{sizeof(size\_t)} expressions because it is now using 16-byte capabilities. This is because the \gls{cheri} architecture invalidates an assumption made by the original \gls{tlsf} developer that \texttt{size\_t} represents the size of a pointer (in memory).
    
\begin{lstlisting}[language=C, caption={An Example Function from TLSF library: offset\_to\_block function. Extract from \cite{tlsf}.}, label={lst:orignal_tlsf}]
static block_header_t* offset_to_block(const void* ptr, size_t size) {
    return tlsf_cast(block_header_t*, tlsf_cast(tlsfptr_t, ptr) + size);
}
\end{lstlisting}

\begin{lstlisting}[language=C, morekeywords={cheri_address_set}, caption={CHERI Ported offset\_to\_block function. Adapted from \cite{tlsf}.}, label={lst:adapted_tlsf}]
static block_header_t* offset_to_block(const void* ptr, size_t size) {
    return tlsf_cast(block_header_t*, cheri_address_set(ptr, tlsf_cast(tlsfptr_t, ptr)+ size));
}
\end{lstlisting}

As illustrated in Listing \ref{lst:orignal_tlsf} and \ref{lst:adapted_tlsf}, to adapt \gls{tlsf} to \gls{cheri}, it is necessary to utilize \gls{cheri} functions to modify capabilities. Indeed, \gls{cheri} use a single-provenance semantics, i.e, every capability needs to be derived from another one. Within this function, acquiring a new capability from an existing one requires the use of \texttt{cheri\_address\_set} to copy the permissions and bounds of \texttt{ptr} to the new capability \texttt{(ptr + size)}.

The porting of \gls{tlsf} to \gls{cheri} required the modification of 7 lines out of a total of 840 lines within \gls{tlsf}, amounting to only 0.83\% of the total codebase.

\subsubsection{Allocator functions}
The next step is to modify the behavior of all the allocation process to manage each memory allocation, and associate a \gls{tlsf}-pool for each domain to obtain a different heap for each one of them. As illustrated in Listing \ref{lst:manager}, each domain possesses its own \gls{tlsf} structure, representing the heap for that particular domain. The initial step involves defining a function to create these separate heaps, as depicted in Listing \ref{lst:heap_init}.

\begin{lstlisting}[caption={Heap creation function.}, label={lst:heap_init}]
void cheri_heap_init(){

    size_t  app_heap_size; 
    uintptr_t  app_heap_address;
    global_manager_s *gm_ptr = &cheri_sdrad_manager;
    domain_info_s *di_ptr = &(gm_ptr->domain_info[gm_ptr->active_domain]);

    char *pTmp;
    pTmp = getenv( "APP_HEAP_SIZE");

    if(pTmp != NULL){
        app_heap_size = atoi(pTmp);
    }else{
        app_heap_size = APP_DEFAULT_HEAP_SIZE;
    }

    app_heap_address = (uintptr_t)mmap(NULL, APP_DEFAULT_HEAP_SIZE, PROT_READ | PROT_WRITE, MAP_PRIVATE | MAP_ANONYMOUS, -1, 0);
    
    if(app_heap_size <= TLSF_MAX_POOL_SIZE){
        di_ptr->tlsf = tlsf_create_with_pool((void *)app_heap_address, app_heap_size);
    }else{
        di_ptr->tlsf = tlsf_create_with_pool((void *)app_heap_address, TLSF_MAX_POOL_SIZE);
        app_heap_size = app_heap_size - TLSF_MAX_POOL_SIZE;
        app_heap_address =  app_heap_address + TLSF_MAX_POOL_SIZE;
        while (app_heap_size  > TLSF_MAX_POOL_SIZE)
        {
            tlsf_add_pool(di_ptr->tlsf, (void *)app_heap_address, TLSF_MAX_POOL_SIZE);
            app_heap_size = app_heap_size - TLSF_MAX_POOL_SIZE;
            app_heap_address = app_heap_address + TLSF_MAX_POOL_SIZE;
        } 
        tlsf_add_pool(di_ptr->tlsf,(void *)app_heap_address, app_heap_size);
    }
}
\end{lstlisting}

The function illustrated in Listing \ref{lst:heap_init} allows the user to create a dedicated heap for the active domain in the application. First, the heap area is allocated with \texttt{mmap()} on line 18. After that, to use the \gls{tlsf} memory allocator, this space needs to be associated with a pool. However, a single call to  \texttt{tlsf\_create\_with\_pool()} or \texttt{tlsf\_add\_pool()} cannot use more than \texttt{TLSF\_MAX\_POOL\_SIZE} for the size of the heap. To address this limitation, a while loop is introduced to add pools until the total size of the heap is reached.

The \texttt{cheri\_heap\_init()} function is not intended to be called directly. Instead, it is invoked by the \texttt{malloc()}-family of allocation functions. This approach optimizes memory allocation by creating a dedicated heap only when necessary. 

To achieve this, the classic \texttt{malloc()}-family of allocation functions were overridden, as illustrated in Listing \ref{lst:Adapted_malloc} with the new version of the \texttt{malloc()} function using the \gls{tlsf} memory allocator.

\begin{lstlisting}[caption={Adapted malloc function.}, label={lst:Adapted_malloc}]
void *malloc(size_t size){
    global_manager_s *gm_ptr = &cheri_sdrad_manager;
    domain_info_s *di_ptr = &(gm_ptr->domain_info[gm_ptr->active_domain]);

    if (di_ptr->heap_init != INIT) {
        cheri_heap_init();
        di_ptr->heap_init = INIT;
    }

    void *ptr;
    size_t rounded_len = __builtin_cheri_round_representable_length(size);

    TLSF_MUTEX_LOCK();
    ptr = tlsf_malloc(di_ptr->tlsf, rounded_len);
    ptr = __builtin_cheri_bounds_set(ptr, rounded_len);
    TLSF_MUTEX_UNLOCK();

    return ptr;
}
\end{lstlisting}

Listing \ref{lst:Adapted_malloc} shows that in this new version of \texttt{malloc()}, the first step is to check if the heap of the active domain is initialized. If it is not, the \texttt{cheri\_heap\_init()} function will be called and the function continues. If it is already initialized, the function just continues. The next step is to round the size of the capability to ensure that it is a multiple of 16. Afterward, the \gls{tlsf} malloc allocator is called instead of the classical \texttt{malloc()} allocator. Finally, the bounds are set to correspond to the size of the capability.
With the same idea, all the allocation functions were overridden to check if the heap is initialized, to use the \gls{tlsf} version of the allocation function, and to correctly set the bounds according to the \gls{cheri} capabilities.

Not only the allocation function needed to be modified, the \texttt{free()} function needed also to be modified. These modifications are illustrated in Listing \ref{lst:new_free}.

\begin{lstlisting}[caption={Modified version of free.}, label={lst:new_free}]
void free(void *ptr)
{
    global_manager_s *gm_ptr = &cheri_sdrad_manager;
    domain_info_s *di_ptr = &(gm_ptr->domain_info[gm_ptr->active_domain]);

    TLSF_MUTEX_LOCK();
    ptr = __builtin_cheri_address_set(di_ptr->tlsf, __builtin_cheri_address_get(ptr));
    tlsf_free(di_ptr->tlsf, ptr);
    TLSF_MUTEX_UNLOCK();
}
\end{lstlisting}

All the capabilities allocated with \gls{tlsf} are associated with a header. This header precedes the capability in memory, specifically before the lower bounds. The approach illustrated in Listing \ref{lst:new_free} involves creating a new capability derived from the entire heap but referencing the address of the capability intended for deallocation. This allows access to the capability's metadata in its header. To achieve this,  \texttt{cheri\_address\_get()} is called to obtain the bounds and permissions of the entire heap using tlsf control structure. Subsequently, \texttt{cheri\_address\_set()} is invoked to generate a new capability using the heap's bounds and permissions, along with the address of the capability planned for deallocation.

\newpage
\section{Evaluation}\label{sec:usecase}
One of the application domains that benefits from increased resilience is service-oriented applications. Servers must have the capability to operate within isolated domains, allowing for the independent shutdown and restart of each domain. Additionally, maintaining a high-level of performance impact is critical for servers. Therefore, the performance of \gls{sdrad} for \gls{cheri} was evaluated on server software to determine if it is a viable approach to enhance security and resilience.

\subsection{Case study: Nginx}\label{sec:nginx}
Nginx \cite{nginx} is a versatile, open-source software that can be used as a web server, reverse proxy, load balancer, and HTTP cache. Originally developed by Igor Sysoev, Nginx is knowed for high performance, stability, and low resource consumption, making it a popular choice for a wide range of web server and proxy server needs.
It is a multiprocessing application featuring a master process and one or more worker processes. The master process oversees the worker processes, which manage client HTTP requests across multiple connections simultaneously. In case of a malicious client request causing memory corruption, a worker process might crash. But fear not! The master process promptly restarts it! Nevertheless, this does mean that any ongoing connections handled by that specific worker are lost in the process.
Moreover, Nginx has recently been ported to \gls{cheri}, making it an ideal fit as a study case for evaluation. The original \gls{sdrad} project \cite{gulmez2023} also used Nginx as a case study.

\subsection{Performance evaluation}
Given its complexity and exposure to untrusted inputs, the HTTP parser stands out as a potential vulnerability within Nginx. The solution involves parsing each client HTTP request within a nested domain. To achieve this, the \texttt{ngx\_http\_parse\_header\_line} and \texttt{ngx\_http\_parse\_request\_line} functions are wrapped using the \gls{api}, as depicted in Listing \ref{lst:wrap_parse_request_line}, specifically focusing on the \texttt{ngx\_http\_parse\_request\_line} function. These functions will be executed instead of the normal ones to proceed with the encapsulation.

\newpage
\begin{lstlisting}[caption={The wrapped ngx\_http\_parse\_request\_line.}, label={lst:wrap_parse_request_line}]
ngx_int_t __real_ngx_http_parse_request_line(ngx_http_request_t *r, ngx_buf_t *b);
ngx_int_t __wrap_ngx_http_parse_request_line(ngx_http_request_t *r, ngx_buf_t *b)
{
    ngx_int_t rc;    

    cheri_domain_enter(NGX_NESTED_DOMAIN); 
    rc = __real_ngx_http_parse_request_line(r, b);
    cheri_domain_exit(); 

    return rc;
}
\end{lstlisting}

Consequently, in the event of detecting memory corruption within the parser, an abnormal domain exit is triggered. This allows us to securely discard the content associated with the nested domain and revert back to the main domain without necessitating a restart of the worker process. Although the connection to the malicious client is closed, all other connections remain unaffected by this operation.

The performance evaluation of Nginx was conducted in four different configurations: baseline (unmodified) Nginx, Nginx ported to \gls{cheri} in purecap mode, purecap Nginx with the \gls{tlsf} allocator and finally purecap Nginx compartmentalized using \gls{cheri}-\gls{sdrad} as described in \cref{sec:nginx}. Throughput measurements were obtained using the WRK \cite{wrk} benchmarking tool. WRK enables us to send a high volume of requests with various configurations to assess the server's performance. This tool allows us to specify the number of simultaneous requests, the number of local machine cores to utilize, and for how long the test should run. It was evaluated with 32 cores, 128 simultaneous requests for 1 minute. The benchmark uses WRK to request files of different sizes (0kB, 1kB, 4kB, 16kB), and to obtain the average result, each benchmark for a specific file size is conducted 10 times to obtain the average result.

The experiment employed two distinct machines: for server deployment, the Arm Morello Board with 4 cores at 2.5GHz and 16GB of RAM running CheriBSD with a FreeBSD \gls{kernel} version 14.0-CURRENT was utilized. Meanwhile, the benchmark was executed on a separate machine equipped with a 32-core Intel(R) Xeon(R) CPU E5-2658 0 clocked at 2.1GHz, along with 66GB of RAM, operating on Ubuntu 22.04.4 with Linux \gls{kernel} 5.15.0. Nginx 1.24.0 release and 1.24.0-with-cheri-fixes release are used, and compiled with -O2 optimization, also to enable Nginx for purecap mode is compiled with \texttt{-target aarch64-unknown-freebsd -march=morello \\-mabi=purecap  -Xclang -morello-vararg=new}.

\begin{figure}[H]
\centering
\includegraphics[width=\columnwidth]{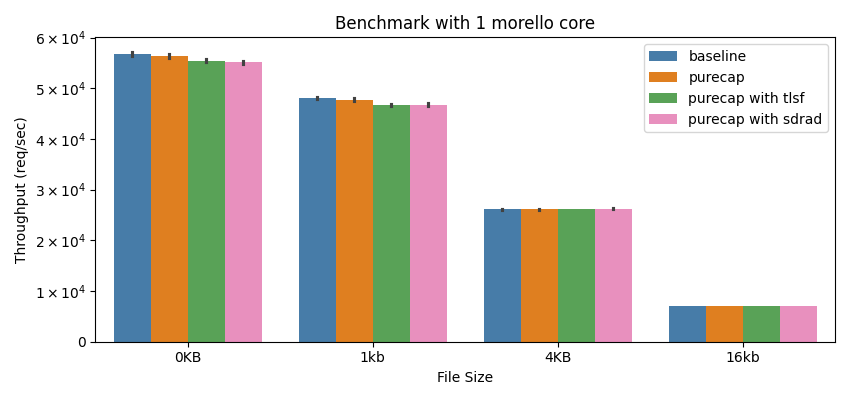}
\caption{Nginx Benchmark with 1 core.}
\label{fig:1_thread_perf}
\end{figure}

Figure \ref{fig:1_thread_perf} summarizes the results of experiments conducted on the Nginx server across four distinct configurations.  Initially, the Nginx server was tested without any additional modifications (baseline), establishing a reference point to observe the impact of different layers of security. The second configuration involved Nginx with the \gls{cheri} modifications running in purecap mode (purecap), serving as the baseline to assess the overhead introduced by the solution. Since \gls{cheri}-\gls{sdrad} requires a specialized allocator, the performance impact of the \gls{tlsf} allocator in \gls{cheri} purecap mode was evaluated without the \gls{cheri}-\gls{sdrad} library (purecap with tlsf). Finally, a version of Nginx in \gls{cheri} purecap mode, compartmentalized with \gls{cheri}-\gls{sdrad} using the \gls{tlsf} allocator (purecap with sdrad), was evaluated.
The result of the benchmark using the \gls{cheri}-port of Nginx using 0kB files, designed to ask the server to send a file of 0kB, showed a 0,78\% degradation of throughput in purecap mode, indicating the performance degradation of \gls{cheri} being negligible compared to unmodified software on the Morello board. Introducing the \gls{tlsf} allocator to the \gls{cheri}-port of Nginx resulted in a slightly higher throughput degradation of 1.73\%, reflecting the increased computational demands. Integrating \gls{cheri}-\gls{sdrad} into \gls{cheri} Nginx led to a 2.22\% throughput degradation, balancing security enhancements with computational efficiency. These results demonstrate the viability of \gls{cheri} integration and highlight the trade-offs between security and performance in server configurations.

\begin{figure}[H]
    \centering
    \includegraphics[width=\textwidth]{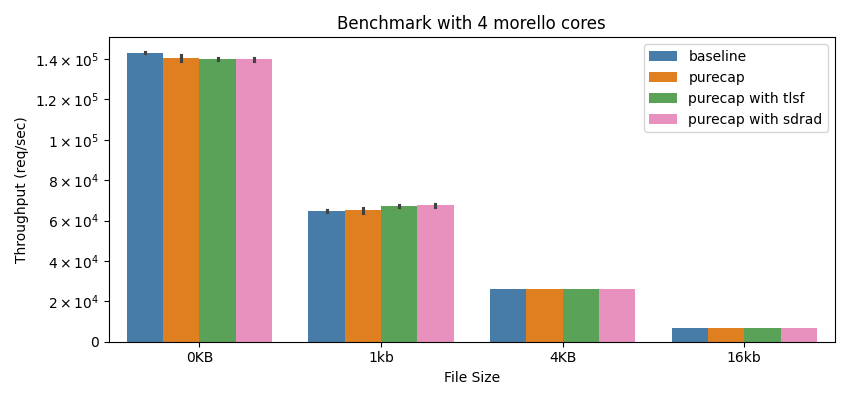}
    \caption{Nginx Benchmark with 4 cores.}
    \label{fig:4_thread_perf}
\end{figure}

Figure \ref{fig:1_thread_perf} illustrates a benchmark conducted using a single core of the Morello Board CPU. Additionally, an attempt was made to utilize all four cores of the CPU, as depicted in Figure \ref{fig:4_thread_perf}. However, the results were not reliable due to the inability to fully saturate the cores of the Morello Board.

\subsection{Comparison to SDRaD on 64-bit x86}
By adapting \gls{sdrad} for the Arm Morello Board, several improvements were noticed:

\begin{enumerate}
    \item During the benchmarking of Nginx with the Intel-based version of \gls{sdrad}, the measured overhead was approximately 5.70\% for 1 worker and 0kb case. Although a direct comparison with the 64-bit x86 version of \gls{sdrad} is not possible, a smaller throughput degradation (2.22\%) was measured for \gls{cheri}-\gls{sdrad} on the Morello board. Additionally, 0.78\% of the throughput degradation was attributed to \gls{cheri} purecap mode, and approximately 0.95\% to \gls{tlsf}, suggesting that the relative impact of \gls{cheri}-\gls{sdrad} on the Morello board is smaller than that of \gls{sdrad} evaluated on an Intel-based architecture. Our conclusion is thus that adapting \gls{sdrad} to a \gls{cheri} architecture resulted in a more lightweight version with less performance impact compared to the baseline 64-bit x64 version architecture.
    
    \item In the 64-bit x64 version \gls{sdrad} a notable drawback of employing Memory Protection Keys (\gls{mpk}) lies in its reliance on tagging memory pages with protection keys using the last 4 unused bits. Consequently, this method is restricted to 16 distinct keys for tagging memory pages. However, when basing compartmentalization on the run-time memory protection capabilities of \gls{cheri}, there is no longer a hardware limitation on the number of domains that can be supported. Although our software implementation limits the number of domains to the number of preallocated domain information storage slots (see \cref{sec:cherisdrad}) this software limitation could be lifted by employing an alternate design that dynamically scales the number of domain information storage slots as needed.
\end{enumerate}

\newpage
\section{Conclusion}\label{sec:conc}

This thesis describes the design and implementation \gls{cheri}-\gls{sdrad} prototype adaption of secure rewind and discard of isolated in-process domains that leverages the memory-safety properties inherent to the \gls{cheri}. \gls{cheri}-\gls{sdrad} results in a design with reduced performance degradation (2.2\% in Nginx benchmarks) compared to earlier results obtained with the original \gls{sdrad} prototype on an Intel-based architecture. The adaption to \gls{cheri} additionally allows limitations inherent to the earlier \gls{mpk}-based approach to be resolved.

The \gls{cheri}-\gls{sdrad} \gls{c} library along with the \gls{cheri}-ported version of \gls{tlsf} memory allocator will be made open source at \url{https://github.com/secure-rewind-and-discard}.

The library developed in this work could also be evaluated for other use-cases, such as Memcached or lighttpd. 
Also, this work can be extended to provide some level of automation similar to \gls{sdrad}-FFI~\cite{sdradffi}.

\newpage
\bibliography{main}

\newpage
\appendix
\section{Appendices}
\subsection{Appendix A: Project Timeline.}
\begin{figure}[H]
    \centering
    \includegraphics[width=0.5255\textwidth]{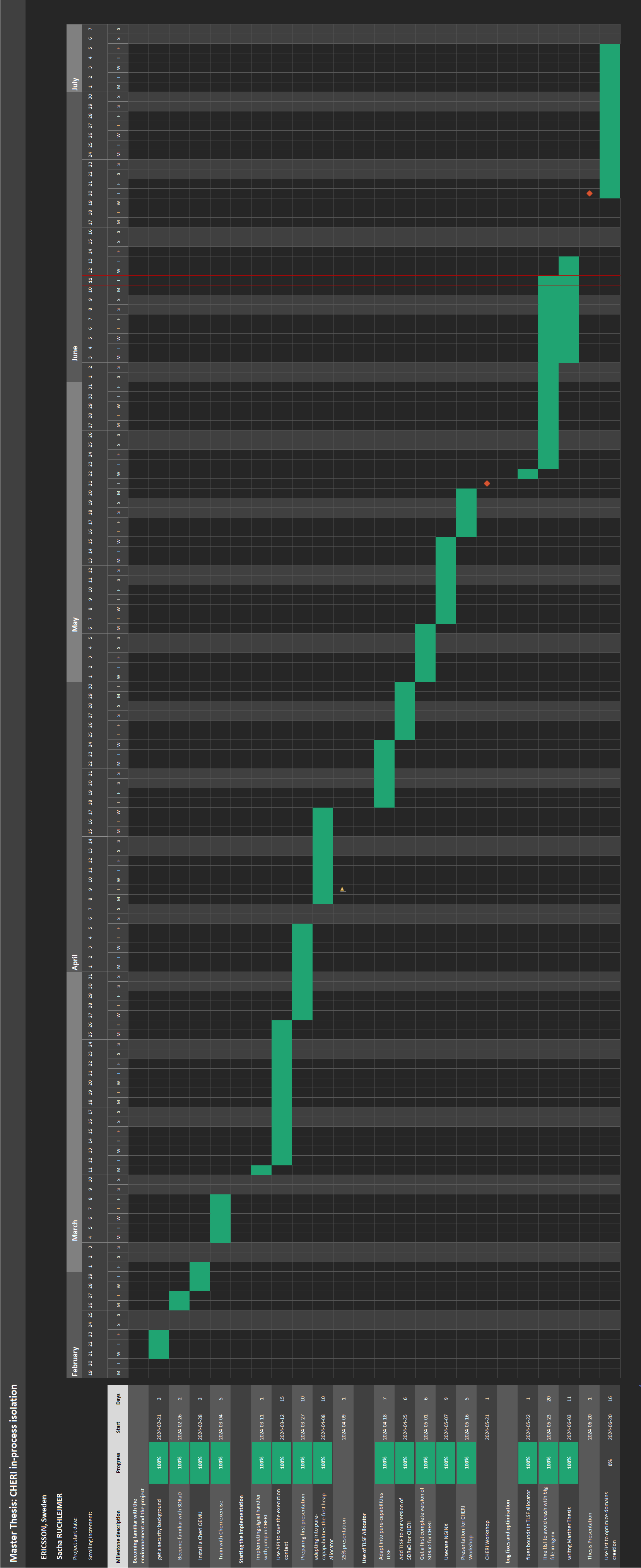}
    \caption{Project Timeline.}
    \label{fig:project_timeline}
\end{figure}

\newpage
\begin{abstract}
\noindent
\begin{minipage}{\textwidth}
Memory-unsafe programming languages such as C and C++ are the preferred languages for systems programming, embedded systems, and performance-critical applications. The widespread use of these languages makes the risk of memory-related attacks very high. There are well-known detection mechanisms, but they do not address software resilience.
An earlier approach proposes the \textit{Secure Domain Rewind and Discard (SDRaD)} of isolated domains as a method to enhance the resilience of software targeted by runtime attacks on x86 architecture, based on hardware-enforced \textit{Memory Protection
Key (MPK)}.
In this work, SDRaD has been adapted to work with the \textit{Capability Hardware Enhanced RISC Instructions (CHERI)} architecture to be more lightweight and performant.
The results obtained in this thesis show that CHERI-SDRaD, the prototype adaption that leverages the memory-safety properties inherent to the CHERI architecture, results in a solution with less performance degradation (2.2\% in Nginx benchmarks) compared to earlier results obtained with the original SDRaD prototype on an Intel-based architecture. The adaption to CHERI additionally allowed limitations inherent to the MPK-based approach to be resolved.
\end{minipage}

\vspace{1em}

\noindent
\begin{minipage}{\textwidth}
\begin{center}
    \textbf{Résumé}
\end{center}

Les langages de programmation non sécurisés en mémoire, tels que C et C++, sont les langages privilégiés pour la programmation système, les systèmes embarqués et les applications nécessitant de hautes performances. L'utilisation répandue de ces langages rend le risque d'attaques liées à la mémoire très élevé. Il existe des mécanismes de détection bien connus, mais ils n'abordent pas la résilience logicielle.
Une approche antérieure propose \textit{Secure Domain Rewind and Discard (SDRaD)} des domaines isolés comme méthode pour améliorer la résilience des logiciels ciblés par des attaques à l'exécution sur l'architecture x86, basée sur la technologie matérielle \textit{Memory Protection Key (MPK)}.
Dans ce travail, SDRaD a été adapté pour fonctionner avec l'architecture \textit{Capability Hardware Enhanced RISC Instructions (CHERI)} afin d'être plus léger et performant.
Les résultats obtenus dans cette thèse montrent que CHERI-SDRaD, le prototype d'adaptation qui exploite les propriétés de sécurité mémoire inhérentes à l'architecture CHERI, offre une solution avec moins de dégradation des performances (2,2\% dans les benchmarks Nginx) par rapport aux résultats antérieurs obtenus avec le prototype original SDRaD sur une architecture basée sur Intel. L'adaptation à CHERI a également permis de résoudre les limitations inhérentes à l'approche basée sur MPK.
\end{minipage}
\end{abstract}

\end{document}

%% file: glossaries.tex
\newacronym{sdrad}{SDRaD}{Secure Domain Rewind and Discard}
\newacronym{cheri}{CHERI}{Capability Hardware Enhanced RISC Instructions}
\newacronym{isa}{ISA}{Instruction-Set Architecture}
\newacronym{api}{API}{Application Programming Interface}
\newacronym{os}{OS}{Operating System}
\newacronym{udi}{UDI}{User Domain Index}
\newacronym{tlsf}{TLSF}{Two-Level Segregated Fit}
\newacronym{mpk}{MPK}{Memory Protection Key}
\newacronym{risc}{RISC}{Reduced Instruction Set Computer}
\newacronym{mmu}{MMU}{Memory Management Unit}
\newacronym{lifo}{LIFO}{Last In, First Out}

\newglossaryentry{c}{
    name={C},
    description={C is a general-purpose, procedural programming language. It provides low-level access to memory and hardware, making it suitable for system programming, such as operating systems and embedded systems}
}

\newglossaryentry{c++}{
    name={C++},
    description={C++ is an extension of the C programming language. It includes object-oriented features such as classes and inheritance, making it suitable for large-scale software development}
}

\newglossaryentry{assembly}{
name={Assembly},
description={Assembly language is a low-level programming language that uses mnemonic codes and symbols to represent instructions that can be directly understood by a computer's CPU (Central Processing Unit). Unlike high-level languages, assembly language corresponds closely to the machine code instructions of a specific computer architecture, making it powerful for tasks requiring precise control over hardware resources and performance optimization},
text={assembly}
}

\newglossaryentry{kernel}{
name = {Kernel},
description= {The kernel is the core component of an operating system, responsible for managing system resources and facilitating communication between hardware and software. It handles critical tasks such as process management, memory management, device management, and system calls, ensuring efficient and secure operation of the computer system.},
text={kernel}
}